\documentstyle[referee]{mn2e}
\input psfig.sty

\def\aa{A\&A \,}

\def\apj{ApJ \,}
\def\aj{AJ \,}
\def\apjl{ApJ Lett. \,}

\def\mnras{MNRAS \,}

\def\kms{km~s$^{-1}$}

\def\Msun{\,M_\odot}
\def\deg{$^o$}
\def\arcsec{$^{\prime\prime}$}
\def\arcmin{$^{\prime}$}

\title[Radio Arcs \& Cosmic Rays in Abell 3376]
{Discovery of giant `radio arcs' in cluster Abell 3376: evidence for shock acceleration 
in a violent
cluster merger?}
\author[Bagchi J.]
{Joydeep Bagchi\thanks{e-mail: joydeep@iucaa.ernet.in}\\
Inter-University Centre for Astronomy and Astrophysics (IUCAA), Post Bag 4, Ganeshkhind, Pune 411007, India \\
}
\date{Accepted ???. Received ????; in original form ????}
\pagerange{\pageref{firstpage}--\pageref{lastpage}}
\pubyear{2002}

\begin{document}
\maketitle
\label{firstpage}
\begin{abstract}

New multi-wavelength (radio, optical \& X-rays) observational 
evidences are presented which show that the nearby ($z=0.046$),
rich cluster of galaxies Abell 3376  is experiencing a
major event of binary subcluster merger. The key evidence 
is the discovery of a pair of  large, optically unidentified     
diffuse radio sources (`arcs'), symmetrically 
located about $\rm 2.6 \, h_{50}^{-1} \ Mpc $
apart at the opposite ends of the hot intra-cluster gas mapped 
by {\it ROSAT} in X-rays. It is argued that the gas-dynamical shock-waves, 
which occur naturally during cluster formation, are accelerating charged particles (cosmic rays)
to relativistic energies, leading to synchrotron emission from
the megaparsec scale radio arcs. If this is so,
cluster Abell 3376 would also be a potential source capable of  
accelerating cosmic ray particles upto ultra-high energies (UHECR) of
$E_{max} \sim 10^{18-19}$ eV. Thus this cluster is an excellent test-bed for
understanding the physics of merger shocks and origin of enigmatic 
UHECR particles in structure formation
process. Hence, Abell 3376   provides unique oppurtunities 
for further multi-wavelength observations with  ground and space-borne observatories.
\end{abstract}


\section{Introduction}\label{intro}

The positive detection of many ultra high energy cosmic 
ray particles  
with surprisingly large energies in the range $10^{18}$ eV to a few 
times $10^{20}$ eV is one of the outstanding enigmas of Physics \cite{Takeda et al. 1999,Hillas}.
It can be shown that if such particles are accelerated 
by some as yet unknown mechanism, the
energy requirements for the astrophysical sources are extraordinary 
\cite{Hillas}. Beyond the `ankle'  of the 
cosmic ray spectrum ($\sim 3\times 10^{18}$ eV),
the Larmor
radius of  UHECR protons in  too large to be retained by the $\sim \mu$G field
of our galaxy 
and therefore they are widely believed to be of extra-galactic origin.
However, beyond the theoretical Greisen-Zatsepin-Kuz'min (GZK) energy cutoff ($\sim 50$ EeV),
which is the proton energy threshold for the photo-pion production on collision with
cosmic microwave background photons (CMB), significant energy losses result. 
In case of $\gamma$-ray primaries, pair creation
through interaction with background photons (CMB, IR and radio waves) is most important
in a wide energy range above the threshold of 
$\sim 4 \times 10^{14}$ eV (cf., review by Nagano \& Watson 2000). Due
to these ubiquitous energy losses, not only it is extremely difficult to accelerate
particles to $\sim 10^{18-21}$ eV, but the  high energy cosmic accelerators 
should probably be found within a horizon of $\sim$ 100 Mpc 
of the observer. So far no astrophysical
cosmic accelerator is visible in the directions from where UHECRs come,
but AGN radio-lobes, AGN central regions, neutron stars, 
gamma-ray burst sources and galaxy clusters,
all have been proposed as the potential sites for UHECR acceleration (Nagano \& Watson 2000 and
references therein). Therefore, identification of nearby accelerator sources
capable of creating highest energy particles is obviously of great importance. 


The largest and the most massive virialized structures are
the galaxy clusters which assemble through the hierarchical process of 
gravitational infall  of smaller mass components. In this process, the accretion of
intergalactic matter and  
collision and merger of smaller groups and clusters take place mainly along 
the axes of large filaments of galaxies 
of length of a few Mpc to up to $\sim$100 Mpc.  
During a merger, the enormous kinetic energy of colliding subclusters 
($\sim 10^{63-64}$ erg) is dissipated in the form of shock-waves which play a pivotal
role in heating of the intra-cluster medium (ICM) to the virial temperature. These Mpc size
super-sonic `merger shock' fronts (of Mach numbers $M \sim 2-5$) have  recently 
been identified in high angular resolution {\it Chandra}
X-ray observations, showing spatial variations of gas temperature and entropy within the 
intra-cluster medium (e.g., Forman et al. 2001). It is also expected that in the peripheral
regions of evolving clusters, there should form large-scale `accretion shocks' due to 
supersonic convergent flows of background  inter-galactic 
plasma \cite{Quilis98,Miniati2000}.

Due to their massive energies, large
sizes and long lifetimes, the shocks, in and around the clusters, are
 potential sites for the acceleration of  very high energy
cosmic-rays (CR) up to $10^{18}- 10^{19}$ eV, near the `ankle' region of CR spectrum
\cite{Norman95,Kang96}, and they may also  be the seeds for 
the ICM magnetic fields \cite{Kulsrud 94}. Recently it has been pointed out
that the cosmic-ray ions accelerated at intergalactic shocks
could accumulate in the formed structure, storing a significant fraction
of the total energy there \cite{mrkj01}. The connection between non-thermal
radiation and cluster mergers has been recently highlighted \cite{Blasi 2001,Sarazin02}. 
Direct evidence for the 
ability of cosmic shock waves to
accelerate particles is given by the observed association of the
so called cluster `radio relic' sources with locations where
shock waves are expected from X-ray observations \cite{Ensslin98}.
Diffusive shock acceleration may be operative at
these locations and responsible for the radio emitting electrons.
In this context, perhaps the strongest evidences are provided by
the discovery of large, diffuse `radio-arcs' in merging cluster Abell 3667
by Rottgering et al. (1997),  and by
the first evidence of shock accelerated relativistic particles and magnetic fields
in a super-cluster scale collapsing filament
of galaxies ZwCl 2341.1+0000 by Bagchi et al. (2002).
Recently, it has been pointed out that some `radio relics' in clusters 
could be the shock wave rejuvenated  fossil remnants
of former active radio galaxies 
\cite{Ensslin01}. 
Although the 
role of shocks in accelerating cosmic-ray 
particles up to $\sim 10^{15}$ eV `knee' energies in  
supernova blast waves is beginning to be understood (e.g., Enomoto et al. 2002),
much remains mysterious about  the astrophysical 
aspects of particle acceleration in large-scale structure formation \cite{Bagchi02}.

This paper  describes the optical, X-ray and radio evidence for an energetic merger  
in A3376 and, reports the discovery of  a pair of giant radio arcs 
(Sect. 2). In Sect. 3, I discuss the
possibility of shock acceleration origin of the radio emission and the implied acceleration of
higher energy cosmic ray particles. The main conclusions 
and the  outlook for future research are summarised
in Sec. 4.
For a cluster redshift of 
0.046, 1 arcmin corresponds to $74.16~h^{-1}_{50}$ kpc, with  the 
Hubble constant expressed in units of 50~km~s$^{-1}$~Mpc$^{-1}$. 

\section{Evidence for merger  in optical, X-ray \& radio wavelengths}

The cluster Abell 3376 (A3376 or DC 0559-40) is an X-ray bright 
cluster, which was  selected for study due to it's remarkable X-ray morphology. It is a member
of the sample of {\it ROSAT} X-ray-brightest Abell-type clusters 
[XBACs; Ebeling et al. 1996. The 0.1-2.4~keV band luminosity is $ \rm
L_{X}= 2.48 \times 10^{44} \, erg \, s^{-1}$],
and also of the {\it Einstein Observatory} cluster sample (Jones \& Forman 1999). This
southern cluster (R.A. $06^h 00^m 43^s$ (J2000),
Dec. $-40^\circ 03^\prime 00^{\prime\prime}$, Bautz-Morgan class I), at the 
redshift z=0.046 \cite{DSa88}, is 
located only $267 \, h_{50}^{-1}$ Mpc away and
 shows several evidences of ongoing energetic merger activity. The
optical galaxies are distributed in a bar-like structure
(projected), extending along a position
angle of $\approx$ 70\deg, defining the merger axis.
The two brightest cluster
members are located near the centers
of  major subcluster of galaxies
which are lined-up along the same position angle 
\cite{DSb88,Escalera94}. 
Fig. 1 shows the optical photograph where  the  
location of two brightest cluster members (shown encircled), 
and other catalogued galaxies (`+' marks; Dressler \& Schectman 1988(a)), have been plotted. 
A strong radio source MRC 0600-399, showing sharply bent
radio jets, is found associated with the bright `E' galaxy positioned at the
{\it ROSAT} X-ray peak (Fig. 1 \& 5). This source also provides important evidence for a merger. 
The relationship between it's radio structure and the 
hydrodynamics generated by merger is further discussed below.

The {\it ASCA} observations
by Markevitch et al. (1998) report a low emission weighted gas temperature (kT$=4.0\pm0.4$
\, keV) and absence of any cooling flow, consistent with the previous findings that
cluster mergers probably disrupt cooling-flows (e.g., Edge, Steward \& Fabian 1992).
The X-ray data from {\it ROSAT} archives further confirms the merger scenario. It reveals a highly
disturbed, non-equilibrium state of the intra-cluster
gas, and the bremsstrahlung X-ray emission elongated along the same
direction as the
optical  axis of the colliding groups (i.e., p.a. $\approx$ 70\deg, 
hereafter called the `merger-axis').
In addition, there are clear evidences for striking 
surface-brightness asymmetry:  i.e., an off-centered `cometary' 
shape, centroid shift, and pronounced
twisting and compression of the inner isophotes near the X-ray peak (see Fig. 2).  
All these 
features are possibly indicative 
of an off-axis merger
scenario \cite{Ricker,Takizawa00}. 

The cross-correlation of {\it ROSAT} X-ray data with  VLA 1.4 GHz NVSS 
atlas \cite{Condon98} reveals 
perhaps the most interesting  aspect of
this cluster: a pair of very large and diffuse radio sources 
(the `radio arcs' hereafter), located
at the opposite ends
of the extended X-ray emitting gas, about 36\arcmin \ or $\rm 2.6 \, h_{50}^{-1} \ Mpc $
apart from each other.
These  features are shown as radio countours (Fig. 1) and as gray-scale image (Fig.
2), overlaid on the optical and X-ray images of A3376. Fig. 3 \& 4 show detailed images,
where the
radio contours  are superposed on the UK Schmidt Telescope
optical images. There is
 no convincing evidence for any  optical
galaxy obviously associated  with the radio arcs, and hence they are 
unlikely to
be the canonical cluster radio galaxies. It is also implausible that they
are radio-lobes of a single giant radio galaxy (GRG), as 
no obvious radio link (jets or plumes) between them and any
central optical galaxy is visible, and GRGs of this large size 
($> 2 h_{50}^{-1}$ Mpc) are extremely rare \cite{Schoenmakers01}. 
On the other hand, the symmetric and
tangential juxtaposition of the 
radio arcs relative to  the merger axis, the brightest cluster galaxies, and  relative to 
the X-ray contours (see Fig. 2)  argue strongly 
that they are part of this cluster and
 very likely originate in
a large-scale energetic process linked to the merger activity. 

\section{Discussion \& Outlook for Future}
Since the  diffuse radio emission is  probably the synchrotron 
radiation, one may ask - what could be the source of  relativistic electrons and 
magnetic fields powering the radio arcs  observed so far ($\approx 1.3 \, h_{50}^{-1}$ Mpc) 
from the cluster center? 
In this respect, the problem is similar
to that of the origin of radio halos, which  require a cluster scale accelerator
source in order to energize electrons distributed over Mpc scales \cite{Jaffe77}.
The radiative life time $t_{IC}$ of an electron with Lorentz factor $\gamma$ in
a weak ($B<3 \ \mu$G) magnetic field is dominated by 
 inverse Compton scattering (IC) on CMB, which
 is, $t_{IC} \approx 2.3 \times 10^{8}  \, \left( {\gamma \over 10^{4}} \right)^{-1} \, 
(1+z)^{-4} \, {\rm yr}$.
The frequency of associated synchrotron emission is $\rm \nu_{syn} \approx 
4.19 \times 10^{8}  \left( {\gamma \over 10^{4}} \right)^{2}  
\left( {B \over \mu G} \right)  Hz$, 
and $\gamma \approx 18300$ (energy $E_{e}=9.35$ GeV) for $\nu_{syn} = 1.4 $ \ GHz, the
 frequency of radio detection.
 
The 3D structure of ICM magnetic fields in merging clusters is 
yet unknown. Nevertheless, according to simulations, it is likely to be 
both significantly tangled on various spatial scales and turbulent \cite{Dolag99,Roettiger99},   
which will severly restrict the diffusion
of charges from their point of injection. In the limit of Bohm type diffusion (diffusion
coefficient $D_{B}(p) \propto$ p, the particle momentum), corresponding
to scattering on saturated field fluctuations, the diffusion length within
the IC cooling time is $l_{diff} \approx (D_{B} \ t_{IC})^{1/2} = 
11.36 \left( {B \over \mu G} \right)^{-1/2}$ pc. Therefore, initial acceleration 
at a central source such as an AGN and diffusive transport upto $\sim$Mpc 
scale is not possible. 
The discrepancy between the 
Bohm diffusion length-scale and the radio structure size is so large 
that, even with inclusion of advective transport by bulk flows 
and more effective diffusion in ordered magnetic fields, 
electrons are still unable 
to cross the emission region within a radiative life-time. It is clear that some form of
{\it in situ} acceleration mechanism is called for. 

Further detailed multi-wavelength observations and 
 computational study of the phenomenon are underway 
in order to understand the details 
of acceleration process involved. Here, it is argued that  
most plausible explanation 
is the diffusive shock acceleration (DSA) \cite{Bell78,BO78} of cosmic
ray particles (electrons, protons and ions) on the  shock fronts, putatively located near
 the radio arcs. 
The prevailing physical
conditions of the disturbed ICM: large-scale bulk flows, 
turbulence, and collisionless MHD shocks, all  provide
ideal environment for particle acceleration through stochastic 
Fermi mechanism \cite{Drury83}, and amplification of seed magnetic fields.

Some important clues to the dynamical history and the energetics of acceleration
process can be obtained from the morphology, flux density,  
and position of the radio arc pair relative to the
cluster center. The  radio peaks   of the western  and eastern 
arcs are located at the positions (J2000) R.A. $06^h 00^m 3^s.09$,
Dec. $-40^\circ 04^\prime 25.0^{\prime\prime}$ (5 mJy/beam, $10\sigma$ detection)
and R.A. $06^h 02^m 59^s.66$,
Dec. $-39^\circ 54^\prime 56.1^{\prime\prime}$ (4 mJy/beam, $8\sigma$ detection),
 and their 1.4 GHz integrated
flux densities are $82\pm5$ mJy and $32\pm3$ mJy respectively. It is noticeable that
both of them are curved, with concave side facing the cluster center, and precisely 
positioned such that a line
joining their radio peaks would run parellel to the merger axis at the position
angle $\approx$ 70\deg . The center of symmetry falls near the center of cluster
 located on the merger
axis at the approximate sky position  R.A. $06^h 01^m 30^s.0$, Dec. $-39^\circ 59^\prime 
50.0^{\prime\prime}$ (shown by a large cross in Fig. 2). The projected 
morphology of radio arcs approximately being the 
arc sectors from a circle of radius $\approx 18^\prime 
\equiv 1.3 \, h_{50}^{-1}$  Mpc, centered
at this position (see Fig. 2). 
In addition, they also
display a distinctive inversion
symmetry -- in the sense that if in the western arc
radio emission extends to the north of
merger axis,  the emission extends to the south of  this axis in case
of the eastern  arc, thereby forming a `S' shaped pattern (Fig. 2). 

These observations suggest a causal relationship between the radio structures 
and the outward propagating shock
waves that probably originated near the cluster 
center $\ga 1.3\times 10^{9}$ {\rm yr} in the past (assuming
constant shock speed of $\sim 1000$ \kms). These shock waves are likely to be   
gravitationally induced  due to observed  merger event of subclusters. 
The integrated radio luminosity of
west(east) arc is $\rm L_{sync} = 9.84(3.84) \times 10^{40}~h^{-2}_{50}~erg \, s^{-1}$ 
(over 10 MHz-100 GHz frequency
range and assumed sp. index $\alpha=-1$). Considering additional IC energy loss, and
a magnetic field $B<3 \ \mu$G, the total radiative losses  add
up to $\rm L_{sync+IC} \approx \, 1.4 \times 10^{42}~ h^{-2}_{50} \, erg \, s^{-1}$.  
Any accelerating process must be energetic enough in order to generate at least this much
amount of non-thermal radiation from electrons. 

One can obtain some order of magnitude estimates by considering the scenario of particle
acceleration in binary cluster merger shocks (e.g., Sarazin 2002).
Numerical simulations
of cluster merger events \cite{Ricker,Takizawa00,MJKR01} 
 demonstrate that these can be very energetic, liberating
gravitational energy  of order $\rm E_{grav} \approx 2.24 \times 10^{63} 
({M \over 2 \times 10^{14} \Msun} )^{2} ({d \over 1.5~Mpc} )^{-1} \, ergs$, 
for two cluster
masses M, at the collision distance d. The gas dynamics is very complex and  
the fluid velocities involved are generally super-sonic -- typically 
$v_{shock} \sim 2000$ \kms ($v_{shock}$ is shock velocity).
The merger  lasts for a duration $t_{merger} \approx d/v_{shock}  \sim
10^{9} \, {\rm yr}$ (which is also the duration of acceleration ). Consequently, 
the  rate of gravitational energy dissipation  is given by
$\rm L_{grav} \approx E_{grav}/t_{merger} \sim 7.1 \times 10^{46} erg \, s^{-1}$. Thus,
typically $\rm L_{grav}$ is much larger than $\rm L_{synch+IC}$ 
and only a  modest fraction of total energy is needed to accelerate the cosmic ray particles. 
The main channel of dissipation of this energy would be the 
shock heating of ICM upto $\rm T_{gas} \sim
10^{7-8} \, K$. Radio observations of supernova shocks with similar $v~\ga~10^{3}$ \kms 
speeds show
that they are capable of converting at least a few percent of shock energy into the energy
of relativistic electrons (e.g., Blandford \& Eichler 1987).


The presence of diffuse radio emissions
ahead of  merging subclusters, marked by bright 
optical galaxies, suggests that they may be
located at the  shock fronts where in situ DSA is a strong possibility. This morphology of
 arcs is  reminiscent of  a classical detached bow shock that preceeds a 
blunt-body in supersonic flight within a fluid.
The elongation of
X-ray contours and the subcluster of galaxies in the same direction (Fig. 2) suggests that
the clusters associated with the bright X-ray core and the cD galaxy are flying
away from each other, possibly $\sim 1$ Gy after their closest encounter.   
However, the `S'
shaped inversion symmetry and curvature  of the radio arcs indicate that the leading bow shocks
have possibly assumed a spiral shape due to an 
off-centered collision 
event (i.e., one with a finite impact parameter, instaed of being head-on), leading to orbital 
motion of coalescing masses about
their common center of mass. It is even possible to infer this sense of rotation from the
radio maps shown, which suggest a clock-wise motion. The pronounced clock-wise
rotation of  isophotes near the {\it ROSAT} X-ray peak, directed towards 
a compressed region on top (Fig. 2), also
suggests this point (see Ricker 1998 and Takizawa 2000 for  numerical simulations of
this geometry). 

From very general principles based on DSA framework, one can derive 
the critical parameters of particle acceleration  in a merging 
system such as A3376. The mean acceleration
time-scale $t_{acc}(E)$ for a particle to reach energy E is determined only by velocity 
jump at the shock and the diffusion coefficients \cite{Drury83}, i.e.,
$t_{acc}(E)~=~\frac{3}{(u_{1}-u_{2})}[(D_1/u_1)+(D_2/u_2)] \approx 
({8/ u_{1}^2})\, D_{B}$. Here $u_{1}(u_{2})$ is the up(down) stream flow velocity and
$D_{1,2}$ are the respective diffusion coefficients. The approximation assumes a strong
shock of compression ratio $r = (u_{1}/u_{2}) = 4$, a frozen-in field condition $(B/\rho)
= const., (D/u)=const.$ across the shock, and Bohm diffusion limit. Under these    
conditions  $t_{acc}= 8.45 \times 10^5 \, u_{3}^{-2} E_{15} B_{\mu}^{-1} Z^{-1}$ {\rm yr}, where
$ E_{15} = (E/10^{15} \, eV)$, $u_{3} = (u_{1}/10^{3}$ \kms), 
$B_{\mu}= (B/10^{-6} \, G)$, and $Z e$ is the nuclear charge. The DSA naturally results in
a power-law for particle energy or momentum function $f(p)$ such that 
$f(p) \propto p^{-b}$, where p is momentum and b is the power-law slope. The synchrotron
spectrum should also be a power-law of the form $I({\nu}) \propto \nu^{-\alpha}$, where
spectral index $\alpha = (b-3)/2$. The index b is related to the compression ratio r
such that $b = 3r/(r-1)$, and $r = 4$, $\alpha = 0.5$ for  
a strong shock in a gas of specific heat 
ratio $\Gamma = 5/3$. Downstream of the `acceleration zone', a `diffusion zone' showing gradually
steepening spectrum is to be expected, which forms due to electrons undergoing diffusion and
advection with the fluid flow and suffering IC and
synchrotron energy losses.
Detailed VLA and GMRT radio spectral and imaging observations of this cluster
are underway in order to test these predictions from the DSA theory. The {\it Chandra}
and {\it XMM-Newton} X-ray telescopes would possibly be able to detect the  temperature and
density jumps associated with the shock fronts. These future 
observations would provide a more definite proof of the  particle acceleration  
scenario proposed for this cluster.

For protons, which suffer negligible radiative losses (below 50 EeV), 
the highest  acceleration energy 
($E_{max}^{P}$) is probably limited by the finite life time of shocks, i.e.,
$t_{acc}=t_{merger}\sim 10^{9-10}$ {\rm yr},  thus giving $E_{max}^{P} \sim 10^{18-19}$ eV. The
heavier nuclei with $Z > 1$ would be accelerated to  even higher energies compared to protons. 
For cosmic ray electrons the situation is different due to 
their significant radiative losses
in synchrotron and IC radiation. The maximum 
energy attained by electrons ($E_{max}^e$) is 
governed by condition that $t_{acc}~\la~ t_{rad} < t_{merger}$, 
and if IC loss dominates over the synchrotron
($B<3 \ \mu$G), one obtains $E_{max}^{e} \sim 3.73 \times 10^{13} \, u_{3} \, 
B_{\mu}^{1/2}$ {\rm eV}. More detailed  
hydrodynamic simulations by Kang, Ryu \& Jones (1996)  show that during gravitational
in-fall and mergers at the sites of evolving clusters, 
energetic protons can be accelerated
upto the GZK  cut-off energy $\sim 50$ EeV, provided a magnetic field of $\sim 1 \,\mu G$
is available to diffusively trap the charge particles near the shock region, and if about
$10^{-4}$ fraction of infalling kinetic energy can be converted as high energy particles.

It is likely that A3376 is a potential source which fits these conditions.
The  magnetic field strength in A3376 cluster is not  known yet,
but could be $\sim 1 \,\mu G$ \cite{Bagchi98,Clarke}. The diffuse
synchrotron emission of radio arcs are `sign-posts' for the existence of relativistic
electrons and magnetic fields  at $\sim$Mpc distances from cluster center.
Since diffusive Fermi-I process accelerates all type of charges (electrons, protons
and ions), it is  argued above that these  structures can in fact be the 
efficient particle accelerators of very energetic cosmic rays.  A3376 is
near enough that some of these cosmic rays could  actually 
be detected by the existing or future
cosmic ray observatories, particularly the energetic 
inverse Compton X-ray and $\gamma$-ray photons
upto extreme energies ${h \nu} \sim 100~{\gamma^{e}_{7}}^{2}$ GeV 
(where ${\gamma^{e}_{7}}$
is the maximum electron Lorentz factor in units of $10^7$). Observation of such particles
would constitute a strong proof of the  cluster shock origin of UHECR particles -- evidence not 
available at the moment.   
On the 
other hand, compared to number of merger events, such 
cluster collision mediated dual shock-fronts appear to be quite rare, because as of today the
only other known example of this phenomenon is a similar radio, optical and X-ray
morphology observed in southern Abell cluster A3667 \cite{Rottgering97}.

The NVSS radio survey has detected the catalogued strong radio source MRC 0600-399 
\cite{molonglo}, associated with the bright 
($m_{pg} \approx 14$) E-galaxy  ($z=0.04552$) positioned at the 
{\it ROSAT} X-ray peak (Figs. 1 \& 2). The higher resolution VLA 4.8 GHz  radio map
shows that both the radio jets are  bent backwards  from the 
radio core in a wide `C' shape (Fig. 5), characterizing it as a wide-angle tail (WAT)
source. WATs are only found in galaxy clusters and  are 
excellent probes of  the
gas-dynamical processes occuring during a merger \cite{Gomez97}. 
Interestingly, here the
northern and  southern radio jets  both appear to be swept back 
along  position angles of $72 \pm 5$\deg ~and $60 \pm 5$\deg. Therefore, even though affected
by projection effects, they appear to be
roughly following the merger axis delineated by optical, radio and X-ray data (Fig. 5).   
This not only indicates correlated  structural distortion, but could also be a good proof of
the plausible hypothesis (e.g., Roettiger et al. 1996) that ram-pressure from bulk gas flows 
of speed $\rm \sim 10^{3} \, km \, s^{-1}$, resulting from mergers, 
are responsible
for bending of radio-jets in WAT  objects in clusters, and not the actual
motion of the parent galaxy with such high speed.  
Thus one can actually use this radio source as
a `wind-sock' to infer which way the cluster wind blows. Hence, more detailed radio and X-ray
observations and modelling of this radio source are likely to prove very instructive.

\section{ Conclusions}
In this paper, I have presented several new observational evidences which show
that the nearby cluster of galaxies A3376 is undergoing a major merger event
of subclusters. The most striking is the discovery of a pair of
optically unidentified, diffuse, giant radio arcs, located $\rm 2.6 \, h_{50}^{-1} \ Mpc $
apart, symmetrically straddling the hot intra-cluster gas imaged by {\it ROSAT}. It is argued
that the likely origin of this Mpc scale radio emission is {\it in situ}
acceleration of relativistic charged particles in gas-dynamical shock-waves, which
occur naturally during cluster formation. Probably the energetics of shocks in A3376  
are strong enough to accelerate cosmic ray particles upto ultra-high energies of
$E_{max} \sim 10^{18-19}$ eV. Thus this cluster is an ideal test-bed for
elucidating the origin of enigmatic UHECR particles in structure formation
shocks. Hence, Abell 3376 is also a target of choice
for more detailed multi-wavelength observations in near future.

\section{Acknowledgments}

Based on photographic data obtained using The UK Schmidt Telescope.
The UK Schmidt Telescope was operated by the Royal Observatory
Edinburgh, with funding from the UK Science and Engineering Research  
Council \& Anglo-Australian Observatory. This work has used
data from the NRAO conducted New VLA All Sky Survey (NVSS) and the
Very Large Array (VLA) telescope.

\vfill
\eject

\begin{figure}
\caption{The VLA 1.4 GHz (NVSS survey) radio contours are shown overlayed
 on the UK Schmidt Telescope  optical image.
The radio beam size (45\arcsec Gaussian) is shown by a circle
(box at lower left corner) and the
contour levels in multiples of 1 mJy/beam  are listed at the bottom. The r.m.s. value of
the background noise is $\approx 0.5$ mJy/beam.
Circles locate the two brightest cluster members and the `+' marks are galaxies
from the  Dressler catalogue (Dressler \& Shectman 1988(a)). The
brightest cD galaxy is inside the circle at lower right. Note the region of missing
radio data towards the south.
}
\protect\label{fig1}
\end{figure}
 
\begin{figure}
\caption{{\it ROSAT} PSPC broad band (0.1-2.4 keV) X-ray image in contours,
overlayed  on the VLA 1.4 GHz NVSS survey radio data shown in
gray scale (Note the missing radio data towards the south). Smaller circles locate the
two brightest cluster galaxies and their subclusters (as in Fig. 1).
To emphasize the
inherent symmetries, a large circle passing through the radio arcs is drawn at the cluster
center (at the large `+' mark), and the merger/symmetry axis is drawn using a dashed
line (see text for details).
}
\protect\label{fig2}
\end{figure}           

\begin{figure}
\caption{VLA NVSS survey 1.4 GHz radio map showing the eastern radio arc.
In the background,
the UK Schmidt Telescope optical image
can be seen. The radio beam size (45\arcsec Gaussian) is shown by a circle and the
contour levels, in multiples of 1 mJy/beam,  are listed at the bottom. The r.m.s. value of
the background noise is $\approx 0.5$ mJy/beam.
}
\protect\label{fig3}
\end{figure}
 
\begin{figure}
\caption{VLA NVSS survey 1.4 GHz radio map showing the western radio arc.
In the background,
the UK Schmidt Telescope optical image
can be seen. The radio beam size (45\arcsec Gaussian) is shown by a circle and the
contour levels, in multiples of 1 mJy/beam,  are listed at the bottom. The r.m.s. value of
the background noise is $\approx 0.5$ mJy/beam.
}

\protect\label{fig4}
\end{figure}     

\begin{figure}
\caption{The VLA 4.8 GHz  (archival data) radio map of the central bent-
jet radio source MRC 0600-399, located at the
 {\it ROSAT} X-ray peak and associated with a bright E-galaxy.
The UKST optical image is shown in the background.
The contour  are drawn at -0.2,-0.1,0.1,0.2,0.4,0.8,1.6,3.2,6.4,12.8 \& 22
mJy/beam. The 24\arcsec Gaussian beam is drawn at the lower left corner.}
\protect\label{fig5}
\end{figure}  

\end{document}